\documentclass[12pt,a4paper]{article}
\pdfoutput=1
\usepackage{epsfig,amssymb,amsmath,graphicx,caption,subcaption,verbatim,
hyperref,xcolor,ulem,epstopdf,psfrag,pstool,braket,array,enumerate,
wrapfig,tabularx,wrapfig}
\usepackage{graphics,color, ltablex,booktabs}
\usepackage[font={small}]{caption}
\numberwithin{equation}{section}

\begin{document} 

\newcommand{\be}{\begin{equation}}
\newcommand{\ee}{\end{equation}}
\newcommand{\bea}{\begin{eqnarray}}
\newcommand{\eea}{\end{eqnarray}}
\newcommand{\bean}{\begin{eqnarray*}}
\newcommand{\eean}{\end{eqnarray*}}
\font\upright=cmu10 scaled\magstep1
\font\sans=cmss12
\newcommand{\ssf}{\sans}
\newcommand{\stroke}{\vrule height8pt width0.4pt depth-0.1pt}
\newcommand{\Z}{\hbox{\upright\rlap{\ssf Z}\kern 2.7pt {\ssf Z}}}
\newcommand{\ZZ}{\Z\hskip -10pt \Z_2}
\newcommand{\C}{{\rlap{\upright\rlap{C}\kern 3.8pt\stroke}\phantom{C}}}
\newcommand{\R}{\hbox{\upright\rlap{I}\kern 1.7pt R}}
\newcommand{\HH}{\hbox{\upright\rlap{I}\kern 1.7pt H}}
\newcommand{\CP}{\hbox{\C{\upright\rlap{I}\kern 1.5pt P}}}
\newcommand{\identity}{{\upright\rlap{1}\kern 2.0pt 1}}
\newcommand{\half}{\frac{1}{2}}
\newcommand{\quart}{\frac{1}{4}} 
\newcommand{\pr}{\partial}
\newcommand{\bm}{\boldmath}
\newcommand{\I}{{\cal I}} 
\newcommand{\M}{{\cal M}}
\newcommand{\N}{{\cal N}}
\newcommand{\e}{\varepsilon}
\newcommand{\ep}{\epsilon}
\newcommand{\balpha}{\mbox{\boldmath $\alpha$}}
\newcommand{\bgamma}{\mbox{\boldmath $\gamma$}}
\newcommand{\blambda}{\mbox{\boldmath $\lambda$}}
\newcommand{\bep}{\mbox{\boldmath $\varepsilon$}}
\newcommand{\Oh}{{\rm O}}
\newcommand{\x}{{\bf x}}
\newcommand{\y}{{\bf y}}
\newcommand{\bR}{{\bf R}}
\newcommand{\bl}{{\bf l}}
\newcommand{\bJ}{{\bf J}}
\newcommand{\X}{{\bf X}}
\newcommand{\Y}{{\bf Y}}
\newcommand{\z}{{\bar z}}
\newcommand{\w}{{\bar w}}
\newcommand{\tT}{{\tilde T}}
\newcommand{\tX}{{\tilde\X}}
\def\ir3{\int_{\mathbb{R}^{3}}}

\thispagestyle{empty}
\vskip 10pt
\begin{center}
{{\bf \LARGE Iterated $\phi^4$ Kinks}} 
\\[20mm]

{\bf \large N.~S. Manton\footnote{email: N.S.Manton@damtp.cam.ac.uk}} \\[1pt]
%\vskip 1em
{\it 
Department of Applied Mathematics and Theoretical Physics,\\
University of Cambridge,
Wilberforce Road, Cambridge CB3 0WA, U.K.}
\vspace{2mm}

{\bf \large K. Ole\'{s} \footnote{email: katarzyna.slawinska@uj.edu.pl}}  \\[1pt]
%\vskip 1em
{\it 
Institute of Physics, \\ Jagiellonian University, Lojasiewicza 11, Krak\'{o}w, Poland\\
}
\vspace{4mm}

{\bf \large A. Wereszczy\'{n}ski  \footnote{email: andrzej.wereszczynski@uj.edu.pl}} \\[1pt]
%\vskip 1em
{\it 
Institute of Physics, \\ Jagiellonian University, Lojasiewicza 11, Krak\'{o}w, Poland\\
}

\vspace{10mm}

\end{center}

%%%%%%%%%%%%%%%%%%%%%%%%%%%%%%%%%%%%%%%%%%%%%%%%%%
\abstract{}
%%%%%%%%%%%%%%%%%%%%%%%%%%%%%%%%%%%%%%%%%%%%%%%%%%

A first order equation for a static $\phi^4$ kink in the presence of
an impurity is extended into an iterative scheme. At the first iteration, 
the solution is the standard kink, but at the second iteration the
kink impurity generates a kink-antikink solution or a bump solution, 
depending on a constant of integration. The third iterate can be a 
kink-antikink-kink solution or a single kink modified by a variant of 
the kink's shape mode. All equations are first order ODEs, so the
$n$th iterate has $n$ moduli, and it is proposed that the moduli space 
could be used to model the dynamics of $n$ kinks and antikinks. 
Curiously, fixed points of the iteration are $\phi^6$ kinks.
  
\vskip 5pt

\vfill
\newpage
\setcounter{page}{1}
\renewcommand{\thefootnote}{\arabic{footnote}}

%%%%%%%%%%%%%%%%%%%%%%%%%%%%%%%%%%%%%%%%%%%%%%%%%%%%%%%%%%%%%%%%%%%%%%%%%%%%%
%%%%%%%%%%%%%%%%%%%%%%%%%%%%%%%%%%%%%%%%%%%%%%%%%%%%%%%%%%%%%%%%%%%%%%%%%%%%%

\section{$\phi^4$ kinks and impurities} 

The $\phi^4$ scalar field theory in one spatial dimension has
Lagrangian
\be
L = \half\int_{-\infty}^\infty \left\{\left(\frac{\pr\phi}{\pr t}\right)^2 
- \left(\frac{\pr\phi}{\pr x}\right)^2 
- (1 - \phi^2)^2 \right\} \, dx
\label{Lagran}
\ee
and dynamical field equation
\be
\frac{\pr^2\phi}{\pr t^2} - \frac{\pr^2\phi}{\pr x^2} 
- 2(1 - \phi^2)\phi = 0 \,.
\label{fieldeq}
\ee
The vacuum solutions are $\phi = \pm 1$ and kinks and antikinks are solutions 
interpolating between these vacua \cite{book,Shn}. The kink satisfies boundary
conditions $\phi \to -1$ as $x \to -\infty$ and $\phi \to 1$ as $x \to
\infty$, and for the antikink the boundary conditions are reversed. Small
and moderate amplitude field oscillations around either vacuum are
interpreted as radiation, and tend to disperse.

As is well known, a static kink obeys the first order
differential equation
\be
\frac{d\phi}{dx} = 1 - \phi^2 \,,
\label{Keq}
\ee
and the family of kink solutions is $\phi(x) = \tanh(x - a)$. The constant of 
integration $a$ is the centre of the kink, and we 
refer to it as the kink's modulus. The manifold of allowed values 
of $a$ (the whole real line) is the moduli space of kink 
solutions. The equation  
\be
\frac{d\phi}{dx} = -(1 - \phi^2)
\label{Kbareq}
\ee
has the antikink solution $-\tanh(x-b)$, and its centre $b$ is the
antikink's modulus.

In kink-antikink dynamics one studies the time-evolution of a 
field that is initially close to a kink centred at $a$ joined to an antikink
centred at $b$, where $b \gg a$. For this configuration, $\phi \to -1$
as $x \to \pm \infty$, but between $a$ and $b$, $\phi$ is initially close to 
$1$. Even at rest, the kink and antikink attract, but the force is
exponentially small in $b-a$. If the kink and antikink are
given initial velocities toward each other, they approach more rapidly. 
The evolution is complicated during the collision. The kink and antikink 
can completely annihilate into radiation (a rather slow process), or 
they can quasi-elastically scatter, emitting less radiation. What 
happens depends sensitively on the initial 
velocities \cite{Sug,Mosh,CSW,Weigel}.

Ideally, one would like to model kink-antikink dynamics in terms 
of a finite number of degrees of freedom, coupled to radiation. To do 
this it is helpful to have a moduli space of field configurations with 
at least two moduli -- one representing the kink-antikink separation, 
and the other the centre of mass. Further to these moduli one can 
consider oscillations of the shapes of the kink and antikink. 
But there is no obvious moduli space available within
the original $\phi^4$ theory. There are no static fields representing
kink and antikink together, because of the attractive force between them. 

One idea is to use the gradient flow curve connecting a well separated
kink-antikink to the vacuum $\phi = -1$. This consists of
the instantaneous field configurations obtained by replacing 
$\frac{\pr^2\phi}{\pr t^2}$ by $\frac{\pr\phi}{\pr t}$ in the
dynamical field equation, and evolving from a well separated kink-antikink 
configuration to the vacuum \cite{Ma4}. These field configurations 
form a moduli space which is fairly closely followed in the true, second 
order dynamics, but the vacuum configuration is an endpoint of this 
moduli space, whereas the true dynamics conserves energy and smoothly 
passes through the vacuum, or close by it, into field configurations 
where $\phi$ is everywhere less than $-1$. The field then continues to 
evolve, oscillating and emitting some radiation in the
process. Gradient flow therefore fails to produce a satisfactory
moduli space in this case.
  
A promising resolution of this difficulty has recently been
identified  \cite{solvable-imp}, based on consideration of the modified static, 
first order equation
\be
\frac{d\phi}{dx} = -(1 - \phi^2)\chi(x) \,.
\label{Impur}
\ee
$\chi$ is referred to as an impurity field, and eq.(\ref{Impur}) as
the kink equation in the presence of an impurity 
\cite{BPS-imp,BPS-imp-susy}. We need to analyse eq.(\ref{Impur}) in 
some detail. Throughout, we assume that $\chi \to -1$ as 
$x \to -\infty$, with the approach sufficiently rapid that the integral
\be
\int_{-\infty}^x (1 + \chi(x')) \, dx'
\ee
converges. We also assume that $\chi \to \pm 1$ as 
$x \to \infty$, and that if $\chi \to -1$ then the integral
\be
\int_{-\infty}^\infty (1 + \chi(x')) \, dx'
\ee
also converges. Only impurities satisfying these conditions
occur in the context of the iterated kinks that will be
introduced in section 2. 

Linearising eq.(\ref{Impur}), we see that $\phi = -1$ is an attractor
as $x \to -\infty$, and $\phi = 1$ a repeller. We can therefore
impose the boundary condition $\phi \to -1$ as $x \to -\infty$, which 
excludes the vacuum solution $\phi(x) = 1$. As $x \to \infty$, 
$\phi = 1$ is an attractor and $\phi = -1$ a repeller in the 
case that $\chi \to -1$, so for generic solutions, $\phi \to 1$ 
as $x \to \infty$. Similarly, $\phi = 1$ is a repeller and $\phi = -1$ an 
attractor in the case that $\chi \to 1$, so $\phi \to -1$ as $x \to \infty$.
Solutions cannot cross $\phi = \pm 1$ so, apart from the vacuum 
solution $\phi(x) = -1$, either $\phi$ is trapped between $-1$ and 
$1$, or $\phi$ is everywhere less than $-1$.

A general impurity field $\chi$ that oscillates between $-1$
and $1$ can make eq.(\ref{Impur}) resemble the original
equations (\ref{Keq}) and (\ref{Kbareq}) in different regions, thus 
allowing for solutions having several kinks and antikinks. We 
stress that these are static solutions of a first order equation.

We can make some more precise statements about the solutions trapped
between $-1$ and $1$ by exploiting Rolle's
theorem. Let us define kink and antikink locations to be precisely
the points $x$ where $\phi(x) = 0$, with $\frac{d\phi}{dx}$ positive
for a kink, and negative for an antikink. The non-generic situation where 
zeros of $\phi$ coalesce and $\frac{d\phi}{dx} = 0$ is where a kink-antikink 
pair is about to be produced or annihilated. Let us focus
on the generic case where $\chi$ and $\phi$ have simple zeros. By Rolle's
theorem, between any pair of distinct zeros of $\phi$ there is a point
where $\frac{d\phi}{dx}$ is zero. Suppose, then, that the impurity
$\chi$ has $N$ zeros. These zeros split the real line into $N+1$
intervals (two of which extend to $\pm\infty$), and there can be at most
one kink or antikink in each of these intervals. $\phi$ therefore has 
at most $N+1$ kinks and antikinks. There can be fewer, by a multiple 
of 2, and the number varies as the constant of integration in the 
solution of eq.(\ref{Impur}) varies. For our choice of
boundary condition they must alternate as kink-antikink-kink-... . 

A simple impurity is the $\phi^4$ kink itself, 
$\chi(x) = \tanh x$, with its zero at the origin. The precise
solution of eq.(\ref{Impur}) for this impurity is given below, but let 
us describe a subset of the solutions more heuristically here. As 
$\tanh x$ is close to $-1$ in the region $x \ll 0$, eq.(\ref{Impur}) resembles 
equation (\ref{Keq}) here, and allows a 
kink solution centred at $-A$, with $A \gg 0$. For $x \gg 0$, $\tanh x$ is 
close to $1$, so eq.(\ref{Impur}) in this region resembles the
sign-reversed equation (\ref{Kbareq}), which admits an antikink
solution centred at $B$, with $B \gg 0$. Solving for
all $x$, one finds a kink-antikink configuration, where the kink is at
$-A$ and the antikink is at $B = A$. The locations are related,
because a first order equation has solutions that depend on only 
one free parameter.

We see that the impurity $\chi(x) = \tanh x$ acts as a mirror. The 
kink part of the solution, around $-A$, is reflected in the
impurity as an antikink around $A$. If the impurity is $\chi(x) = \tanh(x-a)$, 
then there is a solution with a kink at $a-A$ and antikink at $a+A$. There are
now two moduli -- one is the centre of the impurity, and the other
the distance of the kink and antikink from the impurity. We propose 
that the moduli space of these solutions could be used to model the 
kink-antikink fields that occur in the original $\phi^4$ theory 
dynamics. The metric on the moduli space has been calculated 
\cite{solvable-imp}, but there is also a potential energy, that has 
not yet been worked out. Both are needed to define a dynamics on moduli space.

The exact solutions of eq.(\ref{Impur}), for $\chi(x) = \tanh x$, are
\be
\phi(x) = \frac{c - \cosh^2 x}{c + \cosh^2 x} \,.
\label{KKbar}
\ee
The allowed range of the modulus, the constant of integration $c$, is $c >
-1$. Outside this range, $\phi$ has singularities. All solutions satisfy 
the boundary conditions $\phi \to -1$ as $x \to \pm \infty$. The 
kink-antikink configurations, described earlier heuristically, occur for $c$
considerably greater than $1$. Then the zeros of $\phi$ are approximately
where $e^{2x} = 4c$ and $e^{-2x} = 4c$, that is, at 
$x = \pm\half\log(2c)$. These are the locations $\pm A$ of the antikink and
kink. We can check the field profile near $x = -\half\log(2c)$. Just 
keeping the dominant exponential term in $\cosh x$, we find the kink
$\phi(x) \simeq \tanh(x + \half\log(2c))$.

\begin{figure}[h!]
\begin{center}
\includegraphics[width=0.6\textwidth]{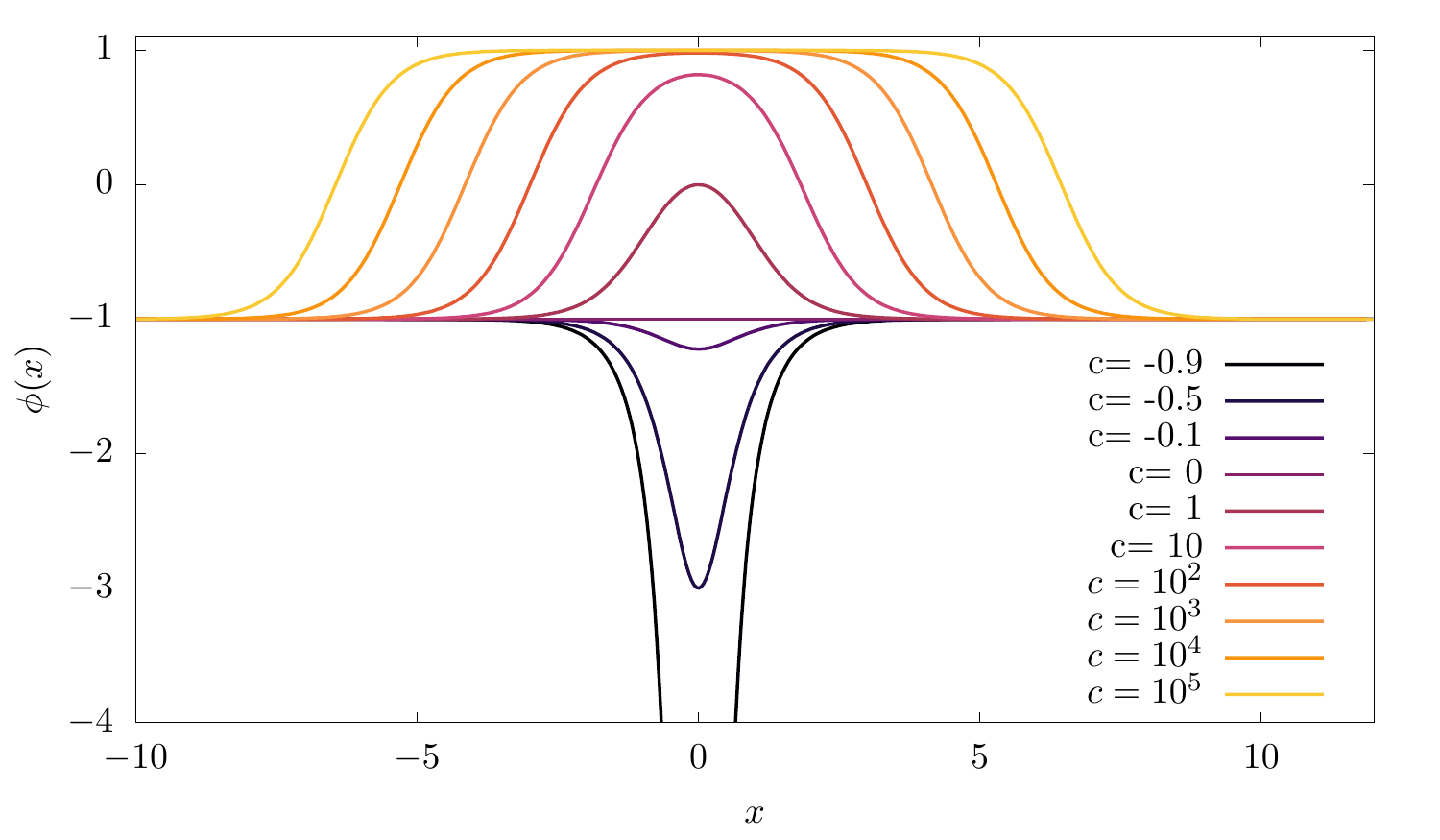}
\caption{Kink-antikink and bump solutions $\phi$.}
%given by eq.(\ref{KKbar}), for values of the 
%modulus $c=10^5, 10^4, 10^3, 10^2, 10, 0, -0.1, -0.5, -0.9$.} 
\label{phi_2}
\end{center}
\end{figure}

When $c = 1$ the kink and antikink annihilate, and for $c < 1$ there
is no kink or antikink, as $\phi$ is nowhere zero. The solution that 
remains we call a bump. For $c$ small, it is a small positive or negative bump 
around $\phi = -1$ of the form 
\be
\phi(x) \simeq -1 + \frac{2c}{\cosh^2 x} \,,
\label{bump}
\ee
and for $c=0$, it reduces to the vacuum $\phi(x) = -1$, 
For $c$ near $-1$ the bump is large and negative, with $\phi \ll
-1$ near the origin. This set of solutions, over the whole allowed range of
$c$, forms a good moduli space for kink-antikink annihilation (with
centre of mass at the origin), better than what is obtained using
gradient flow, because it interpolates from well separated kink and 
antikink, through the vacuum, to a large negative bump. See Fig. \ref{phi_2}. 
All these configurations occur in kink-antikink dynamics.

Another impurity that has been considered in \cite{BPS-imp-phi4} 
is of the bump shape (\ref{bump}), 
\be
\chi(x) = -1 + \frac{2c}{\cosh^2 x} \,,
\label{BumpImpur}
\ee
with $c$ not necessarily small. For $c=0$, one solution of 
eq.(\ref{Impur}) is the standard kink centred at the origin, but for 
$c$ small and non-zero, the kink becomes deformed by a variant of
the shape mode \cite{Ra1}. For $c > \half$ the impurity (\ref{BumpImpur}) has 
two zeros. This allows the kink to be sufficiently deformed that it 
becomes a kink-antikink-kink configuration.

Recall that the shape mode is a small,
normalisable deformation of the kink with frequency of oscillation 
$\omega = \sqrt{3}$ according to the linearised dynamical equation 
(\ref{fieldeq}) for $\phi$. The continuum of radiation modes have 
frequencies $\omega \ge 2$, and the kink's translation zero mode has
frequency $\omega = 0$. A kink distorted by both a zero mode of amplitude
$\alpha$ and a shape mode of amplitude $\beta$ has the form
\be
\phi(x) = \tanh x + \alpha\frac{1}{\cosh^2 x} +
\beta\frac{\sinh x}{\cosh^2 x} \,.
\label{Kinkdeform1}
\ee

For the impurity (\ref{BumpImpur}), with $c$ small, there are solutions of 
eq.(\ref{Impur}) close to the standard kink that are similar to 
(\ref{Kinkdeform1}). To see this, set $\phi(x) = \tanh x + \eta(x)$ and 
work to linear order in both $\eta$ and $c$. Eq.(\ref{Impur}) then reduces to
\be
\frac{d\eta}{dx} = -2\tanh x \, \eta - \frac{2c}{\cosh^4 x} \,,
\ee
and this linear inhomogeneous equation has the general solution
\be
\eta(x) = \alpha\frac{1}{\cosh^2 x} - 2c\frac{\sinh x}{\cosh^3 x} \,,
\label{Kinkdeform2}
\ee
combining a zero mode of arbitrary amplitude with a modified 
shape mode where the power of $\cosh x$ in the denominator is $3$ not $2$.

This is interesting. The shape mode usually arises through oscillations
of the kink, but here a variant arises independently through the effect of a
small-amplitude bump impurity which itself arises (approximately) 
as a solution of the
kink equation with kink impurity (\ref{Impur}). This hints at an exact 
iterative scheme that could capture more of the degrees of freedom needed to 
model kink-antikink dynamics using a finite-dimensional moduli space. 
It has long been recognised that an effective model for kink-antikink 
dynamics should allow not only for the kink-antikink separation, but also 
for the shapes of the kink and antikink to be deformed 
\cite{Sug,CSW,Weigel,TW}. The shape mode also plays a role in 
(symmetric) kink-antikink-kink dynamics \cite{MM}. A 
kink-antikink-kink configuration can annihilate into a single kink, 
emitting radiation, and the approach towards annihilation is 
approximately tangent to the shape mode of the surviving kink.    

All this suggests that useful moduli spaces of multiple kink-antikink
configurations can be found as exact solutions of an iterated kink
equation with impurity. We describe this next.

\section{Iterated kinks}

Our proposed iterated kink equation is
\be
\frac{d\phi_n}{dx} = -(1 - \phi_n^2)\phi_{n-1} \,, \quad n =
1,2,3,\dots \,,
\label{Iter}
\ee 
where we fix $\phi_0(x) = -1$. 

We impose the boundary condition $\phi_n \to -1$ as $x \to -\infty$, 
for all $n$, and also require that $\phi_n$ has no singularities. 
This allows the vacuum solution
$\phi_n(x) = -1$ for any $n$, but excludes $\phi_n(x) = 1$. The 
boundary condition appears to be consistent, by the following inductive
argument. Obviously $\phi_0$ satisfies the boundary condition, and 
linearisation of eq.(\ref{Iter}) about $\phi_n = -1$ shows that if
$\phi_{n-1}$ satisfies the boundary condition, then $\phi_n(x) \sim -1 +
\mu e^{2x}$ for $x \ll 0$ and some constant $\mu$, and hence $\phi_n$ 
satisfies the boundary condition.   

The iteration can go on indefinitely, introducing one extra modulus
each time. On the other hand, for each $n$ there is always the vacuum
solution $\phi_n(x) = -1$, whatever the form of $\phi_{n-1}$, 
and one can iterate this repeatedly and get the vacuum for all 
larger $n$. The iteration has then effectively stopped at the $(n-1)$th step.

Iterating the argument in section 1 concerning the attractive and
repulsive natures of $\phi = -1$ and $\phi = 1$, we deduce that for
generic solutions of eq.(\ref{Iter}), $\phi_n \to 1 \, (-1)$ as 
$x \to \infty$ for $n$ odd (even). Exceptionally, the sign may be 
reversed if one or more fields $\phi_k$ in the solution sequence is the 
vacuum, $\phi_k(x) = -1$.

Equation (\ref{Iter}) for $\phi_n$ is simply the kink equation 
(\ref{Impur}) with impurity $\phi_{n-1}$, and as each equation in the sequence
is first order, its solution has one constant of integration. Iterating, 
and allowing these constants to be free, we may interpret $\phi_n$ 
as having $n$ moduli. The arguments in section 1 concerning zeros 
of $\phi$ imply that $\phi_n$ has at most $n$ zeros, and if it has 
the maximal number, it is interpreted as a solution with $n$ kinks 
and antikinks, whose locations are a choice for the moduli.
In this case, the iterated kink equation adds one new kink or
antikink to the solution at each step. This is reminiscent of a
B\"acklund transformation in sine-Gordon theory, although the details
seem quite different. 

The first few iterates are field configurations we have previously
discussed. $\phi_1$ obeys the standard kink equation (\ref{Keq}), 
having solution $\phi_1(x) = \tanh(x-a)$ with arbitrary centre $a$. 
Notice that the equation also has the solutions $\phi_1(x) = -1$ 
and $\phi_1(x) = \coth(x-a)$ satisfying the boundary condition, but 
the latter is excluded because it is singular at $x=a$.

For the second iteration, let us take $\phi_1$ to be the kink centred at 
the origin, as the effect of a translation is rather trivial. Equation
(\ref{Iter}) for $\phi_2$ is the same as equation (\ref{Impur}) with impurity 
$\tanh x$, having the solutions (\ref{KKbar}) illustrated in Fig. \ref{phi_2}. 
For all $x$ and $c$, $\phi_2(x) < 1$. As before, the solutions include 
kink-antikink pairs, and also positive and negative bumps on
the background of the vacuum $\phi_2(x) = -1$.

Also acceptable at the second iteration is for the impurity to be the
vacuum, $\phi_1(x) = -1$. This gives solutions for $\phi_2$ that are
either simple kinks or again the vacuum. The family 
of solutions $\phi_2$ therefore incorporates all acceptable
solutions in the $\phi_1$ family, including the starting, vacuum solution
$\phi_0$. An interpretation is that the family of generic $\phi_2$ kink-antikink
solutions is completed by sending the antikink to infinity, and then
both the kink and antikink to infinity.

\begin{figure}[h!]
\includegraphics[width=0.5\textwidth]{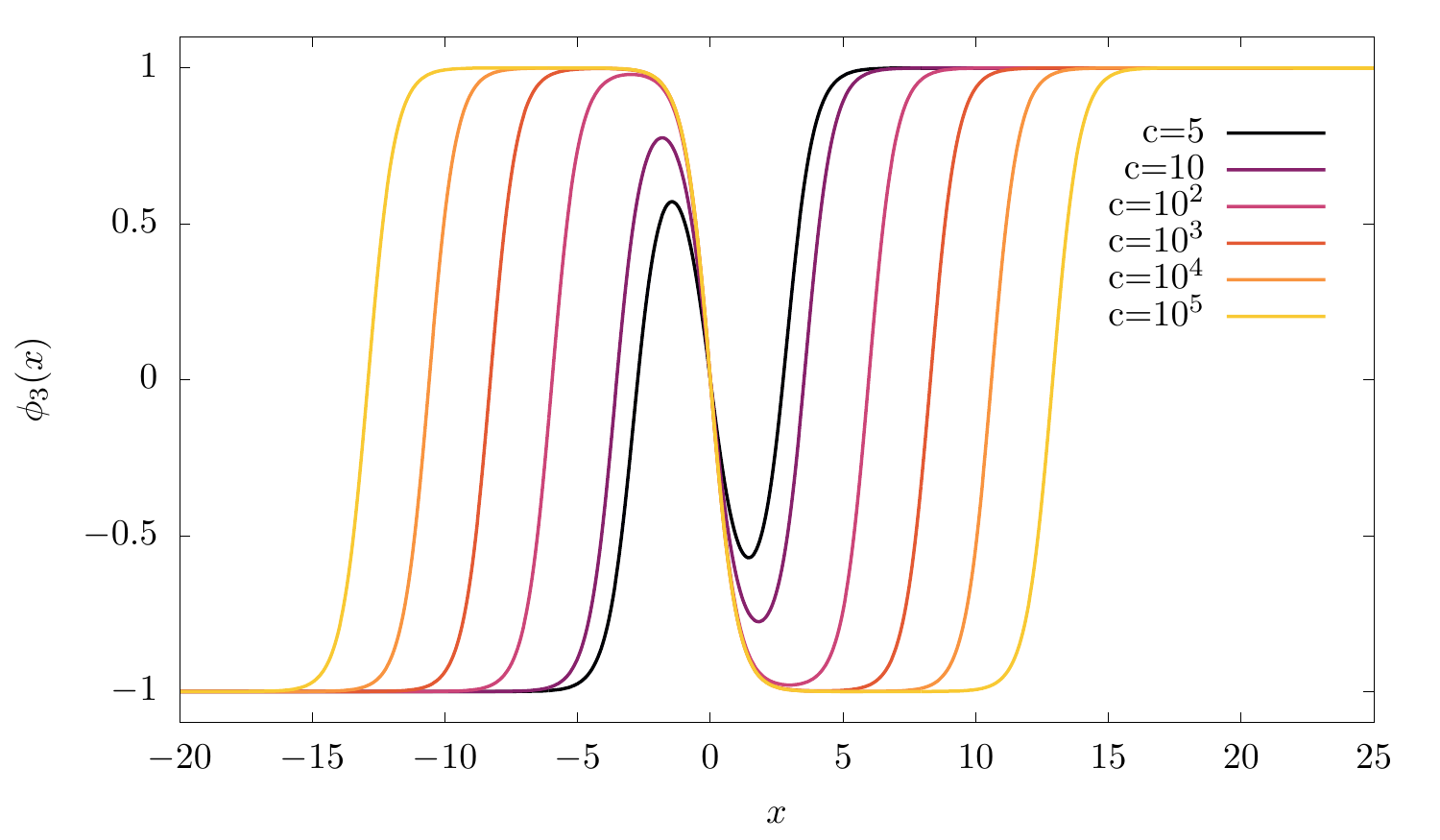} 
\includegraphics[width=0.5\textwidth]{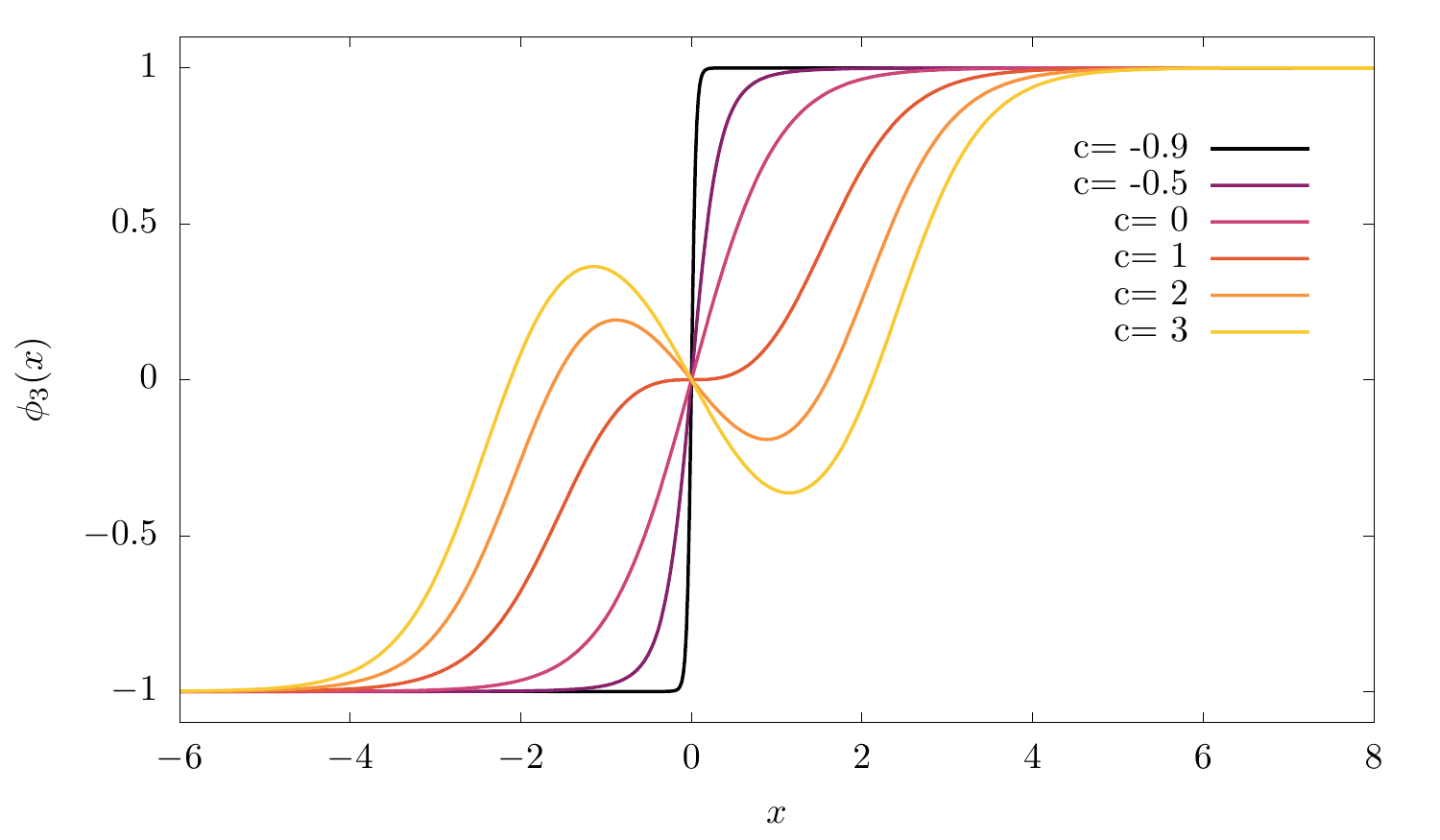} 
\caption{Reflection-symmetric solutions $\phi_3$ for various 
values of $c$.} \label{phi_3_a}
\end{figure}
\begin{figure}[h!]
\includegraphics[width=0.5\textwidth]{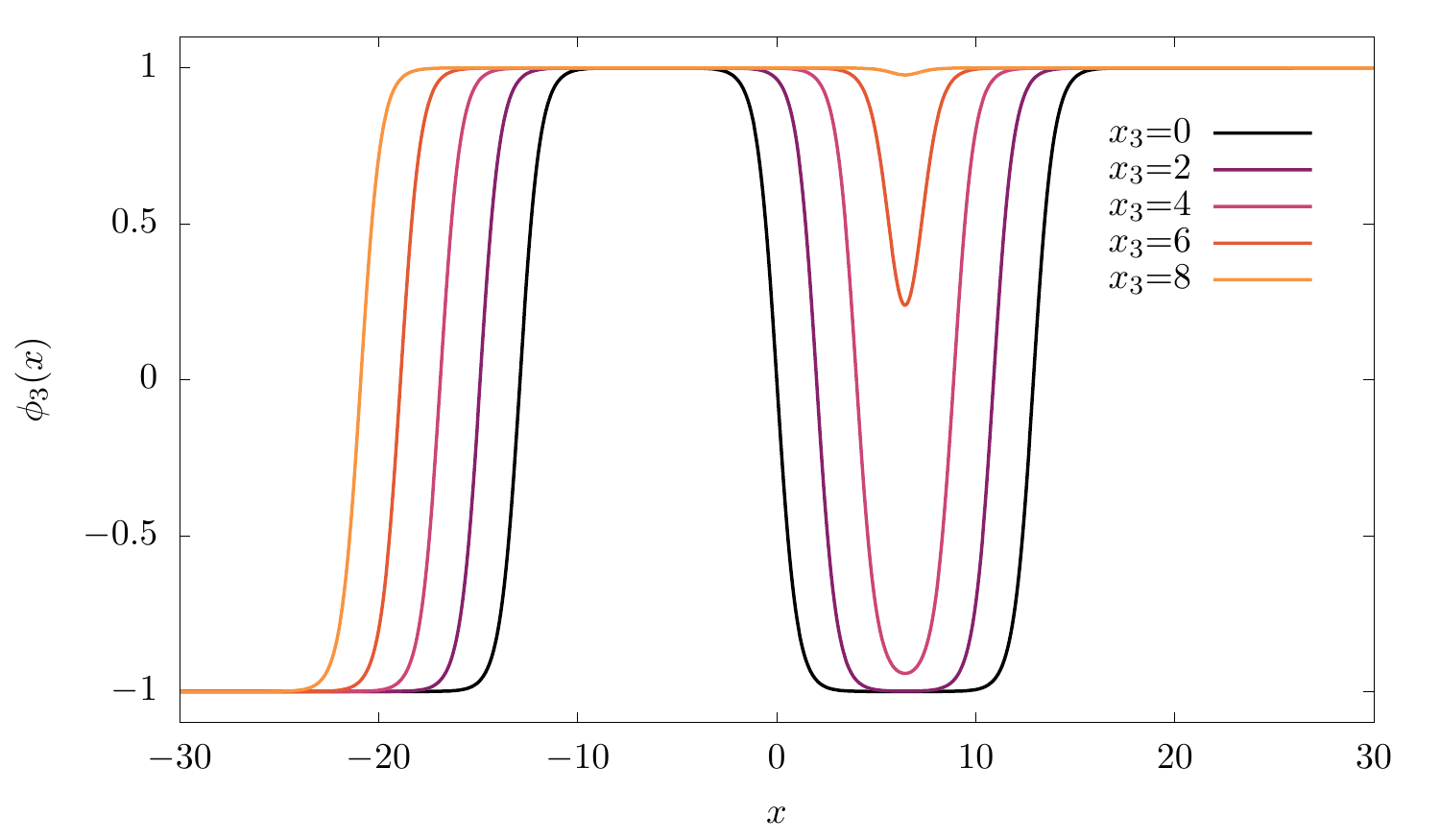} 
\includegraphics[width=0.5\textwidth]{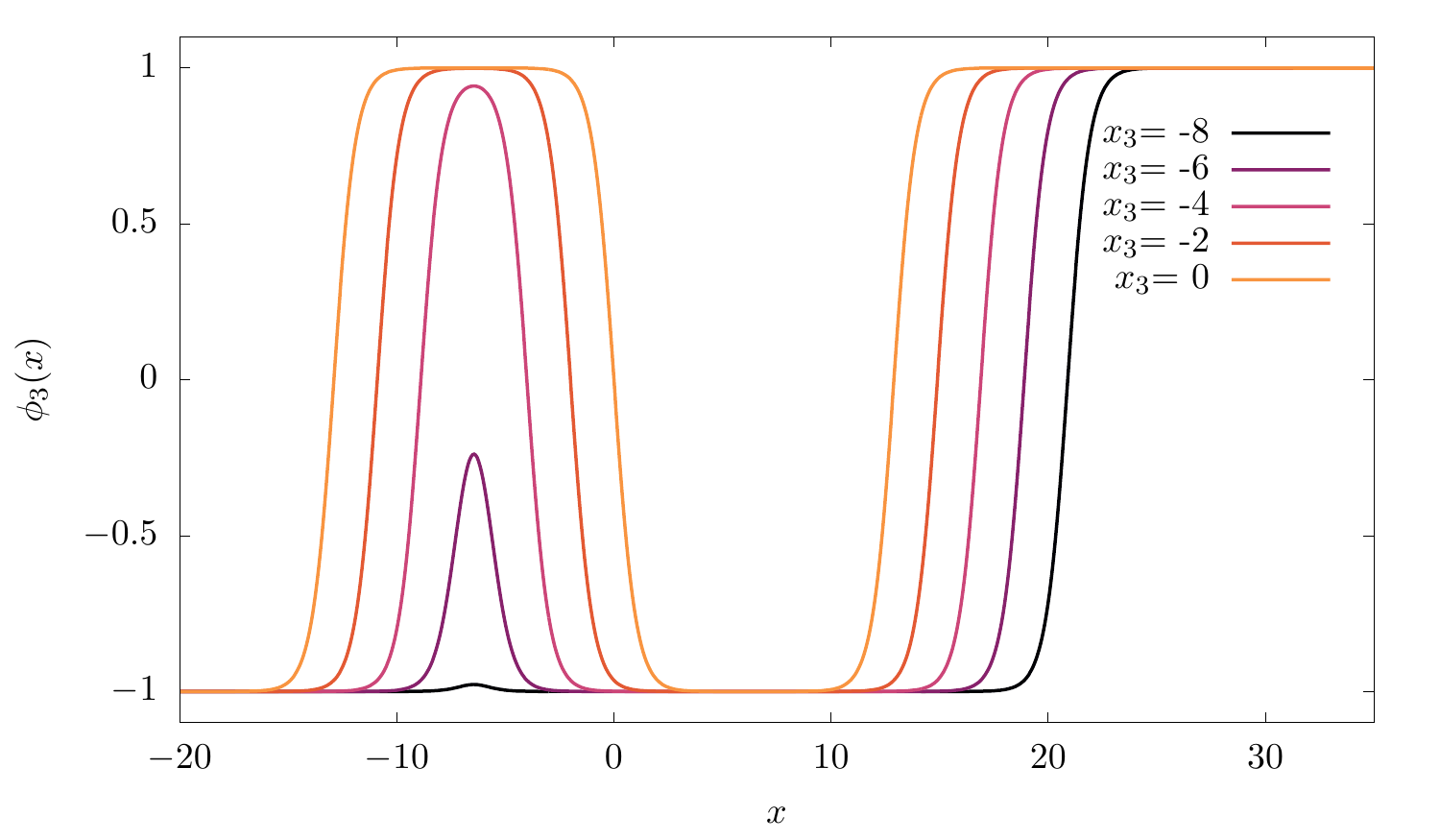} 
\caption{Reflection-asymmetric solutions $\phi_3$ for $c=10^5$.} 
\label{phi_3_b}
\end{figure}
\begin{figure}[h!]
\includegraphics[width=0.5\textwidth]{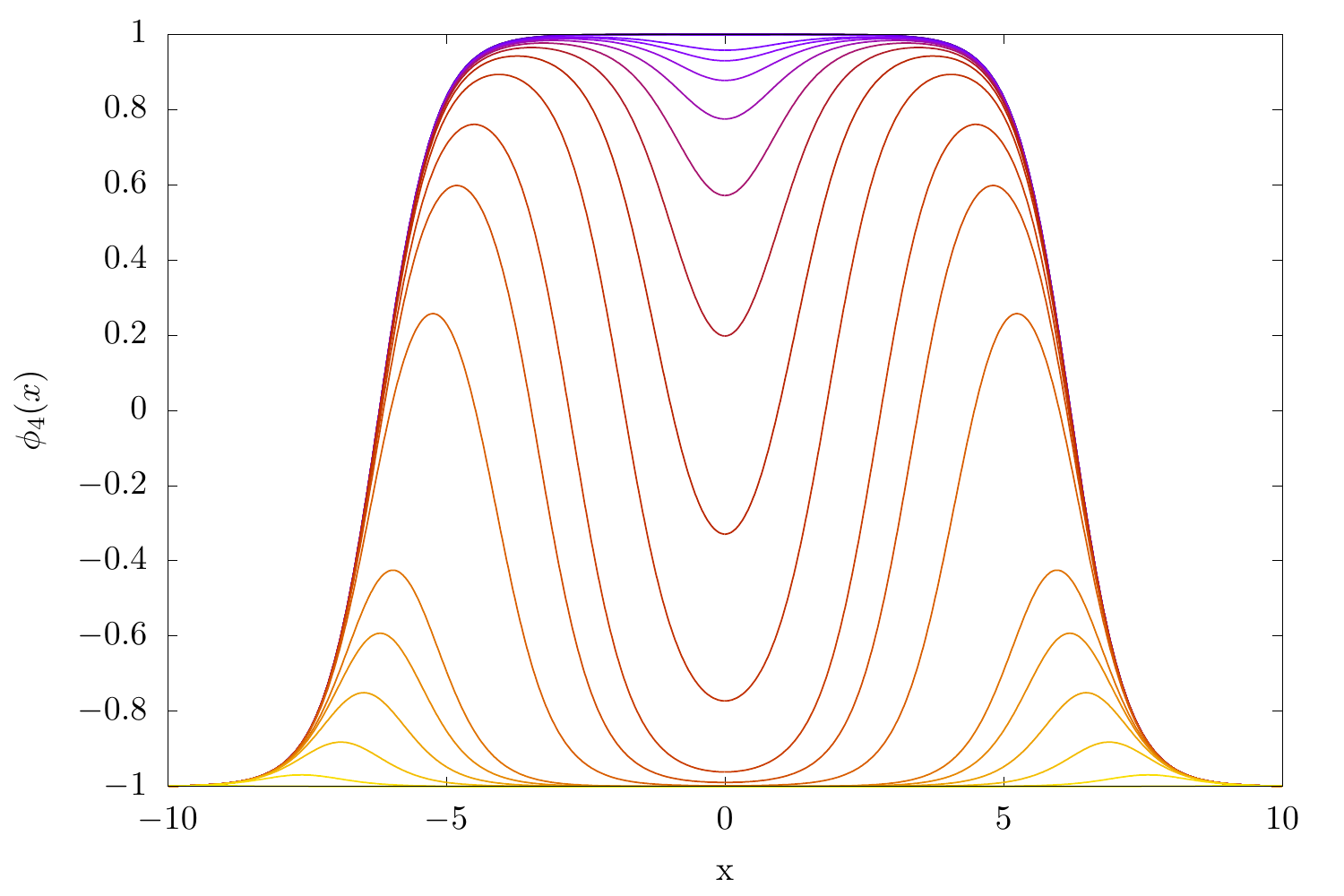} 
\includegraphics[width=0.5\textwidth]{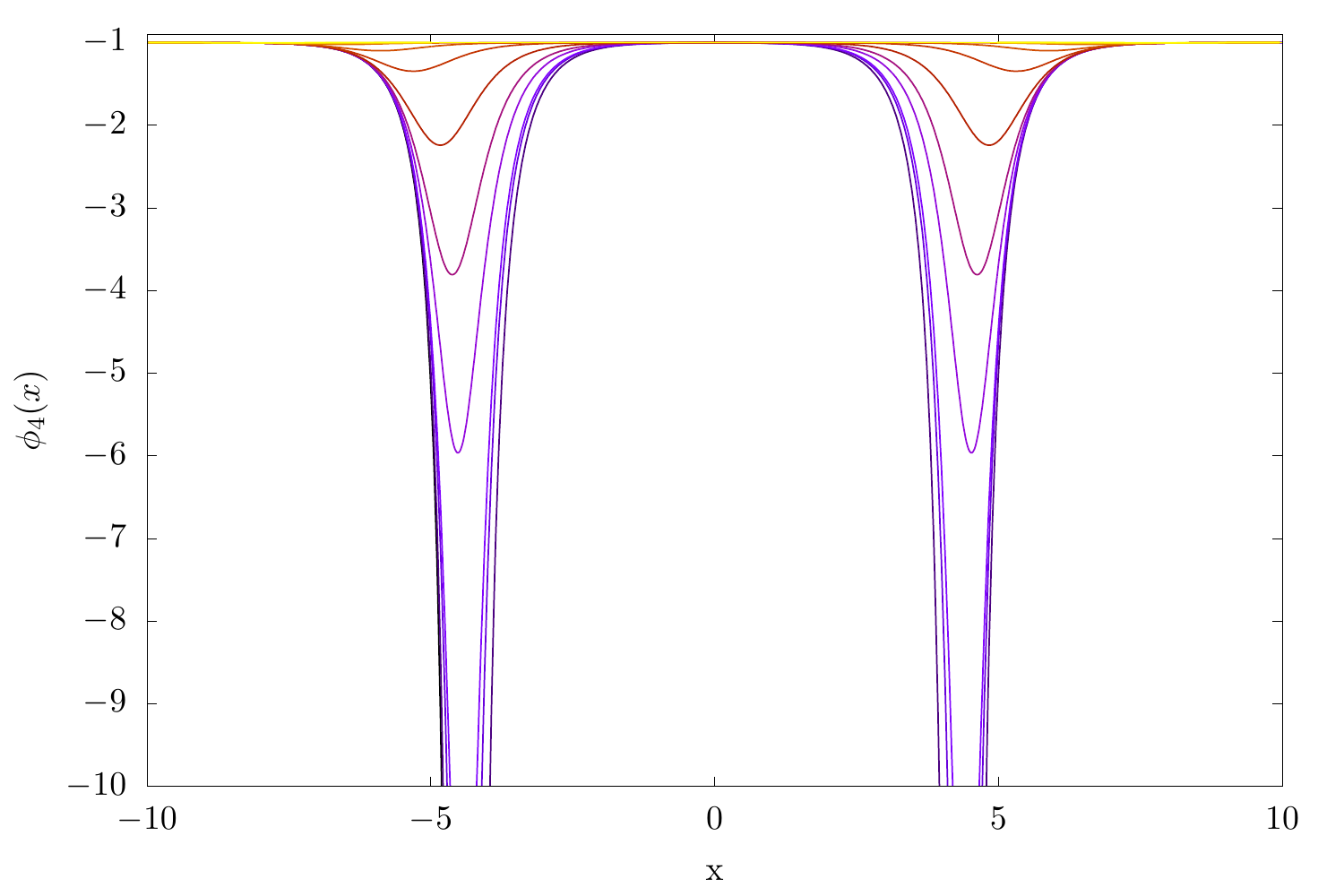} 
\caption{Examples of $\phi_4$ solutions.} \label{phi_4}
\end{figure}

The third iteration is algebraically more complicated. We need to 
solve eq.(\ref{Iter}) for $\phi_3$ with impurity $\phi_2$ given by 
eq.(\ref{KKbar}). The explicit solution is given in section 3. The 
three moduli of the solution are the constant of integration $x_3$, the 
parameter $c$ in $\phi_2$, and the centre of the original kink 
$\phi_1$. Particularly interesting are the solutions with the 
reflection symmetry $\phi_3(-x) = -\phi_3(x)$, which arise when $\phi_1$ is 
a kink at the origin, $\phi_2$ has arbitrary parameter $c > -1$, and 
the constant of integration is chosen to preserve the symmetry. These 
solutions are shown in Fig. \ref{phi_3_a}. Use of their 1-dimensional 
moduli space could resolve some difficulties in modelling 
kink-antikink-kink dynamics that arose in ref. \cite{MM}. Note the 
appearance of a shape deformation when $c$ is close to zero, as we 
anticipated in the approximate solutions (\ref{Kinkdeform2}), and 
the occurrence of kink-antikink-kink solutions for $c > 1$.
Fig. \ref{phi_3_b} shows a class of solutions $\phi_3$ without
reflection symmetry, with fixed $c = 10^5$ and various constants of
integration $x_3$  (see eqs.(\ref{phi3tanh}) and (\ref{phi3tan})).

We have not systematically attempted a fourth iteration but can make 
some general observations. A class of solutions $\phi_4$
consists of kink-antikink-kink-antikink configurations. If these are 
well separated we can denote their locations, where $\phi_4(x) = 0$, 
by $a_1, a_2, a_3, a_4$. The kink-antikink 
pair at $a_1$ and $a_2$ arises from a kink impurity at their midpoint
$\half(a_1 + a_2)$. Similarly the antikink-kink pair
at $a_2$ and $a_3$ arises from an antikink impurity at
$\half(a_2 + a_3)$, and so on. So $\phi_3$ is a 
solution with kink, antikink and kink locations $\half(a_1 + a_2), 
\half(a_2 + a_3), \half(a_3 + a_4)$. In turn, $\phi_3$ arises from 
a kink-antikink solution $\phi_2$
with kink and antikink locations $\quart(a_1 + 2a_2 + a_3), 
\quart(a_2 + 2a_3 + a_4)$, and finally $\phi_2$ arises from a single 
kink impurity $\phi_1$ centred at $\frac{1}{8}(a_1 + 3a_2 + 3a_3 + a_4)$.

Not all solutions $\phi_4$ are well separated kink-antikink-kink-antikink 
configurations. Some such solutions, and some alternative types of
solution involving bumps, are shown in Fig. \ref{phi_4}. There could also be an
interesting class of solutions $\phi_6$ with two moduli. These would be 
configurations with a reflection symmetry, where a kink on the left is
deformed, and an antikink on the right is similarly 
deformed. The moduli space could be similar to that proposed in 
\cite{Sug} and further discussed in \cite{Weigel, TW}.   

\section{Space-deformed kinks}

The equation for a kink with impurity (\ref{Impur}) can be formally
integrated \cite{solvable-imp}, and this solution method gives considerable
geometrical insight. The method can be applied 
iteratively to solve the entire set of equations (\ref{Iter}),
but the result involves multiple integrations, and appears
algebraically intractible.

Recall that the right hand side of (\ref{Impur}) vanishes for 
$\phi = \pm 1$, so solutions cannot cross these values. A 
solution $\phi(x)$ that approaches $-1$ as $x \to -\infty$ is 
either (i) trapped between $-1$ and $1$, or (ii) is everywhere
less than $-1$. We ignore here the vacuum solution $\phi(x) = -1$.

Let us first rewrite eq.(\ref{Impur}) as
\be
\frac{d\phi}{1 - \phi^2} = -\chi(x) \, dx \,.
\label{Impur1}
\ee
In case (i), the solution is
\be
\tanh^{-1}\phi \equiv \half\log\left(\frac{1 + \phi}{1 - \phi}\right) 
= -\int^x \chi(x') \, dx' \,,
\label{intsoln1}
\ee
where the lower limit of the integral provides a constant of
integration. In case (ii), the solution is
\be
\coth^{-1}\phi \equiv \half\log\left(\frac{\phi + 1}{\phi - 1}\right) 
= -\int^x \chi(x') \, dx' \,.
\label{intsoln2}
\ee
(Note that the argument of the logarithm is positive in both cases.)
The solutions exploiting the hyperbolic functions are more familiar in
the context of kinks, but the logarithmic form can be easier to
manipulate algebraically.

The right hand sides of both
(\ref{intsoln1}) and (\ref{intsoln2}) can be expressed as
\be
-\int^x \chi(x') \, dx' = x - a - \int_{-\infty}^x (1 + \chi(x')) \,
dx' \,,  
\ee
with $a$ arbitrary. (Recall that we are assuming that the last
integral converges.) We refer to 
\be
y(x) = x - \int_{-\infty}^x (1 + \chi(x')) \, dx'
\ee
as the deformed spatial coordinate. The solutions (\ref{intsoln1}) and
(\ref{intsoln2}) are then simply $\phi(x) = \tanh(y(x) - a)$ and 
$\phi(x) = \coth(y(x) - a)$. These clearly satisfy eq.(\ref{Impur1}), 
since $dy = -\chi(x) dx$.

As $y$ is finite for all (finite) $x$, the solution $\tanh(y(x) - a)$ is
always acceptable. The solution $\coth(y(x) - a)$ is acceptable only if $y$
remains less than $a$ for all $x$, otherwise there is a singularity.
For example, for $\chi(x) = \tanh x$, $y(x) = -\log(2\cosh x)$, which
has a maximal value of $-\log 2$. So the $\coth$ solution is acceptable
only for $a > -\log 2$. The tanh and coth solutions together reproduce
the solutions (\ref{KKbar}). 

The interpretation of the solution $\tanh(y(x) - a)$ depends on the
behaviour of $y$ as $x$ increases. If $\chi$ is everywhere negative, 
which means that $\chi \to -1$ as $x \to \infty$, then $y$ increases 
to $\infty$ monotonically with $x$, and the solution is a spatially 
deformed single kink. If $\chi < -1$ everywhere, then 
$y$ increases more rapidly than $x$. The effect is to produce a 
solution $\phi(x)$ that is a steepened kink. If $\chi$ 
crosses zero at $x = X$, then $\frac{dy}{dx}$ changes sign and part of the 
profile of $\phi$ is reflected about $X$. Equivalently, there is
a spatial fold at $X$. If $\chi$ crosses zero again, there is another 
reflection, or fold. 

When $\chi \to -1$ as $x \to \infty$, we can define the
overall stretching or compression of the deformed kink,
\be
s = \int_{-\infty}^\infty (1 + \chi(x')) \, dx' \,.
\ee
The asymptotic form of $\phi(x)$ is $\tanh(x-a)$ for $x \ll 0$ and 
$\tanh(x-a-s)$ for $x \gg 0$. The kink has been stretched by distance 
$s$ if $s>0$ and compressed by $|s|$ if $s<0$. Stretching by more than 
a small distance can introduce kink-antikink pairs.

Analogous to the spatial folding in the relation between $y$ and
$x$ is to imagine walking the length of a corridor, when it is
uncomfortable to walk very slowly, but comfortable to sit for a 
while. One can walk the length in one go (a kink), and sit the rest 
of the time, or walk backwards and forwards a few times (kinks and 
antikinks), sitting less. With more time available one can walk more 
often backwards and forwards. If the time available is short, one must 
walk quickly (a steepened kink). 

All this analysis applies to the iterated kink equation. Consider a
generic sequence of solutions $\phi_n(x)$. For $n$ odd, $\phi_n$ must
be of the $\tanh$ type, to avoid singularities, but for $n$ even, 
$\phi_n$ can be of $\tanh$ or $\coth$ type. For $n$
odd, $\phi_n$ is a spatially deformed kink,
\be
\phi_n(x) = \tanh \left(x - x_n 
- \int_{-\infty}^x (1 + \phi_{n-1}(x')) \, dx' \right) \,,
\ee
whose deformed spatial coordinate is
\be
y_n(x) = x - \int_{-\infty}^x (1 + \phi_{n-1}(x')) \, dx' \,,
\label{yn}
\ee
and whose overall stretching/compression is
\be
s_n =  \int_{-\infty}^\infty (1 + \phi_{n-1}(x')) \, dx' \,.
\ee
$x_n$ is the arbitrary constant of integration.

An explicit solution for $\phi_3$ can be found using this
approach. Let us assume that $\phi_1$ is a kink centred at the origin;
$\phi_2$ is then given by eq.(\ref{KKbar}). Using the deformed spatial
coordinate $y_3$ given by the integral (\ref{yn}), we obtain for $c
\ge 0$,
\be
\phi_3(x) = \tanh\left(x - x_3 - \frac{2c}{\sqrt{c(1+c)}}
\tanh^{-1}\left(\sqrt{\frac{c}{1+c}}\tanh x\right)\right) \,,
\label{phi3tanh}
\ee
and for $-1 < c \le 0$,
\be
\phi_3(x) = \tanh\left(x - x_3 - \frac{2c}{\sqrt{-c(1+c)}}
\tan^{-1}\left(\sqrt{\frac{-c}{1+c}}\tanh x\right)\right) \,.
\label{phi3tan}
\ee
The solutions are shown in Fig. \ref{phi_3_a} and Fig. \ref{phi_3_b}. 
Specifically, in Fig. \ref{phi_3_a} we plot $\phi_3$ for $x_3 = 0$. 
These are the solutions with reflection symmetry. It is clear
that the modulus $c$, which measures the strength of $\phi_2$,
controls the emergence of an antikink-kink pair. For large $c$ such a 
pair is easily visible in $\phi_2$, and the whole solution $\phi_3$ 
represents a kink-antikink-kink configuration. When $c$ approaches 
zero, $\phi_2$ tends to the constant $-1$, which leads to a single kink for
$\phi_3$. This single kink solution becomes steeper and steeper as $c
\rightarrow -1$. 

In Fig. \ref{phi_3_b} we show the impact of $x_3$ on $\phi_3$ 
for fixed $c$. We choose $c=10^5$ to better visualise the observed 
behaviour. Here, $\phi_2$ represents a well separated kink-antikink 
pair. For large $x_3$ the solution describes a single kink
monotonically interpolating between the vacua. The impact of $\phi_2$ 
is negligible, except on part of the kink tail. When $
x_3$ approaches zero, the single kink interacts 
strongly with $\phi_2$ and the kink-antikink pair hidden in $\phi_2$ 
has a pronounced effect. Finally, for large negative $x_3$, the 
single kink reappears but on the opposite side of the origin. 
This variation with $x_3$ represents a flow on the moduli space,
where an incoming kink creates an antikink-kink pair (due to the
interaction with $\phi_2$), and later on annihilates this pair 
leaving an outgoing kink.

\section{$\phi^6$ kink as fixed point}

The iterated kink equation has a curious fixed point. We find this by
setting $\phi_n = \phi_{n-1}$. Then eq.(\ref{Iter}) becomes the
$\phi^6$ kink equation
\be
\frac{d\phi}{dx} = -(1 - \phi^2)\phi \,.
\label{phi6}
\ee 
The generic non-singular solutions, satisfying the boundary condition 
$\phi \to -1$ as $x \to -\infty$, are of the form
\be
\phi(x) = -\left( 1 + 2e^{2(x-c)} \right)^{-\half} \,,
\label{phi6kink}
\ee
with $c$ arbitrary. These all have the property $\phi \to 0$ 
as $x \to \infty$.

We have not constructed an iterated sequence of solutions $\phi_n$
with limiting form (\ref{phi6kink}). The approach to the limit cannot
be uniform in $x$.

There is also an interesting 2-cycle of the iteration, a
solution of the pair of equations
\bea
\frac{d\phi}{dx} &=& -(1-\phi^2)\psi \,, \\
\frac{d\psi}{dx} &=& -(1-\psi^2)\phi \,.
\eea
We assume that $\phi \to -1$ and $\psi \to -1$ as $x \to -\infty$.
Setting $\psi = \phi \Omega$, we find that these equations reduce to
\bea
\frac{d\phi}{dx} &=& -(1-\phi^2)\phi\Omega \,, \label{phiOmeg} \\
\frac{d\Omega}{dx} &=& -(1-\Omega^2) \label{Omeg} \,.
\eea
The equation (\ref{Omeg}) for $\Omega$ is the usual first order 
equation for a $\phi^4$ antikink. It has trivial solutions 
$\Omega(x) = \pm 1$, and non-trivial solutions 
$\Omega(x) = -\tanh x$ and $\Omega(x) = -\coth x$, or translates of these.
If $\Omega(x) = 1$ then we recover the fixed point solution (\ref{phi6kink}), 
the $\phi^6$ kink. There is no solution with $\Omega(x) = -1$ satisfying the
boundary conditions. When $\Omega(x) = -\tanh x$, then 
$\psi(x) = -\phi(x)\tanh x$, so $\phi(x) = -\psi(x)\coth x$. Therefore, 
multiplication by $-\tanh x$ and $-\coth x$ automatically alternate 
during iteration of the 2-cycle, so we need only consider the 
case $\Omega(x) = -\tanh x$.  

\begin{figure}%[h!]
\includegraphics[width=0.5\textwidth]{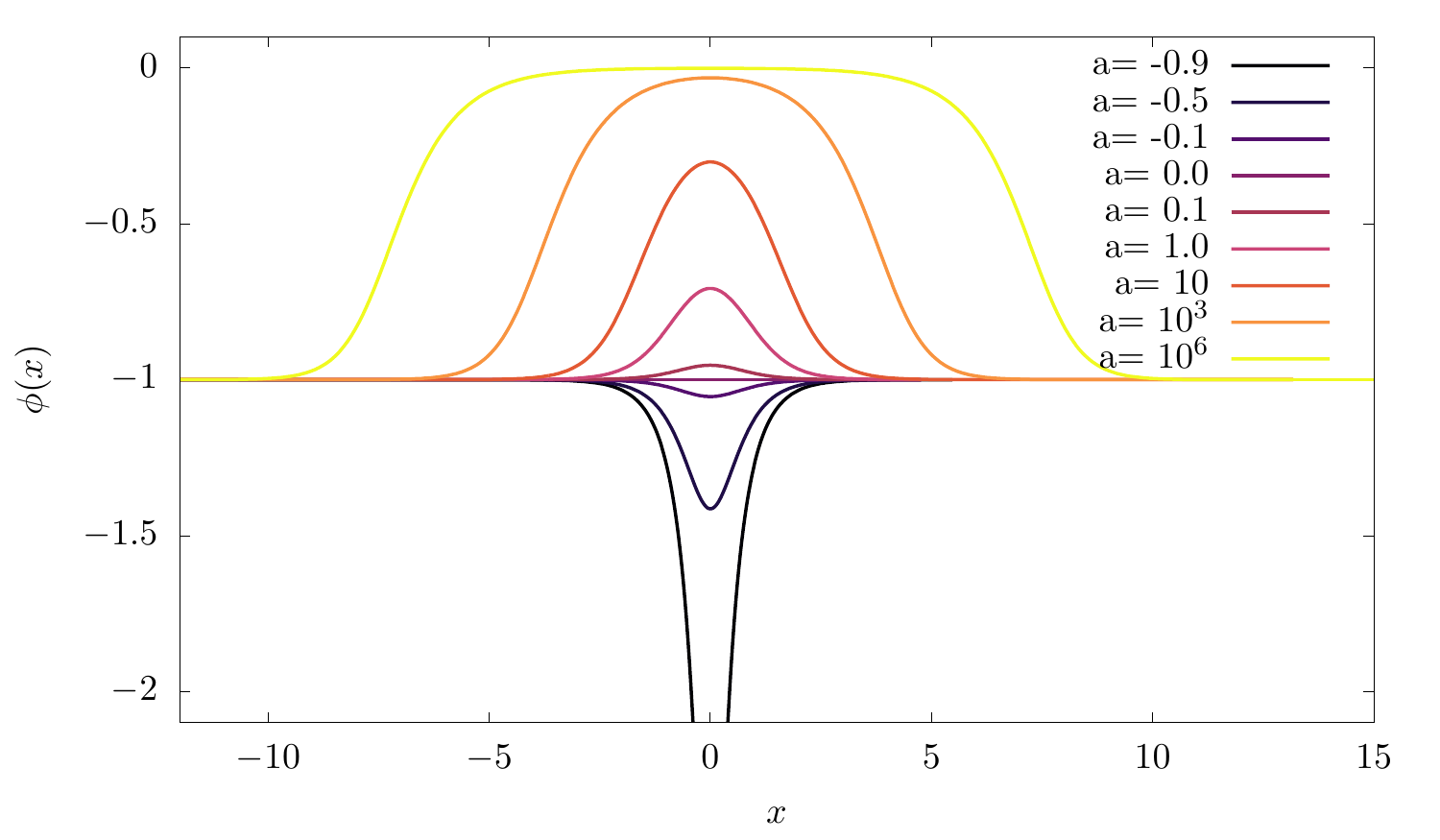}
\includegraphics[width=0.5\textwidth]{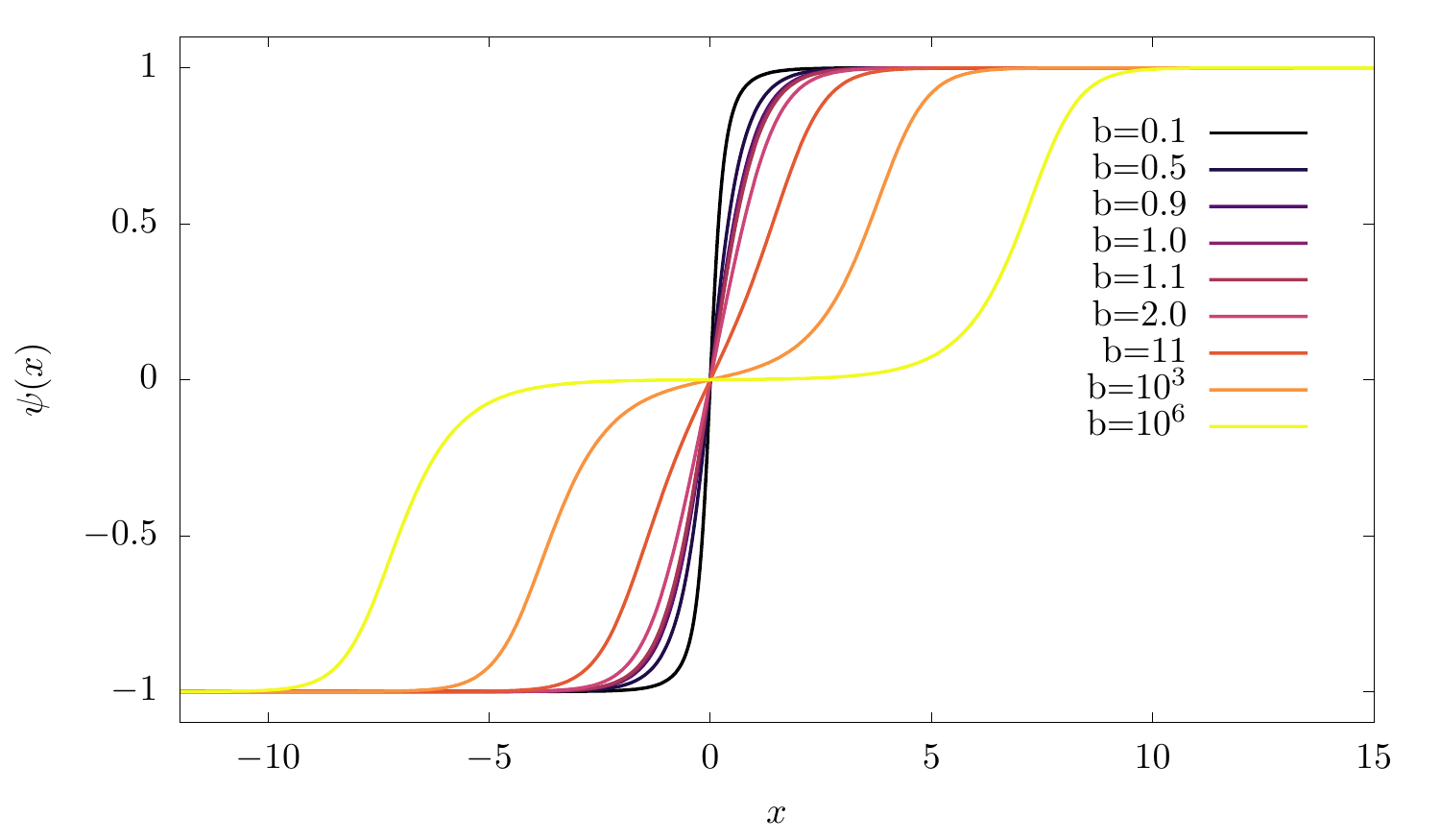} 
\caption{2-cycle solutions $\phi$ (left) and $\psi$ (right) 
for various values of $a$ and $b$.} \label{phi6figs}
\end{figure}

The remaining equation (\ref{phiOmeg}) can be expressed as
\be
\frac{d\phi}{(1-\phi^2)\phi} = \tanh x \, dx \,.
\ee
This is the standard equation for a $\phi^6$ kink, but in terms of a
deformed spatial coordinate $y$, defined by $dy = \tanh x \, dx$. 
Integrating, we find that $\phi(y) = \pm(1 + 2e^{-2(y-c)})^{-\half}$, where 
$y = \log \cosh x$. Choosing the appropriate sign, and rearranging, 
we obtain the solution
\be
\phi(x) = -\left(1 + \frac{a}{\cosh^2 x}\right)^{-\half} \,,
\label{2cyclephi}
\ee
with $a > -1$, and this is paired in the 2-cycle with
\be
\psi(x) =\mbox{sign} (x) \left(1 + \frac{b}{\sinh^2 x}\right)^{-\half} \,,
\label{2cyclepsi}
\ee 
where $b = a + 1$. See Fig. \ref{phi6figs}. For large positive $a$ the field
$\phi$ describes a well separated kink-antikink pair of
$\phi^6$ theory (interpolating between $-1$ and $0$). When $a$
decreases the kink and antikink approach each other and finally, for $a=0$,
annihilate to the vacuum $\phi = -1$. For negative $a$ the field $\phi$
forms a negative bump whose strength becomes arbitrarily large as $a$ approaches
$-1$. Simultaneously, the field $\psi$ represents a kink (interpolating
between $-1$ and $0$) and a second kink (interpolating between $0$ and 
$1$) of $\phi^6$ theory. They separate completely as $a \rightarrow
\infty$. When $a=0$ these kinks merge into the kink of $\phi^4$ theory,
and this becomes steeper and steeper as $a \rightarrow -1$.
 
\section{Energy function for iterated kinks}

Here we present an energy function whose stationary points include the
solutions of the iterated kink sequence of equations (\ref{Iter}). 
Let us start with eq.(\ref{Impur}) for a kink with given impurity $\chi$, 
rewritten as
\be
-\frac{1}{\chi(x)}\frac{d\phi}{dx} = 1 - \phi^2 \,.
\label{Impur2}
\ee
We shall suppose that the zeros of $\chi$ (if
any) are a discrete set of points, and require that 
$\frac{d\phi}{dx}$ is zero at these points. Recall that 
$\phi(-\infty) = -1$, and $\phi(\infty) = \pm 1$. 

Consider the energy
\be
E_{\chi} = -\half\int_{-\infty}^\infty 
\left\{ \frac{1}{\chi^2(x)}\left(\frac{d\phi}{dx}\right)^2 
+ (1 - \phi^2)^2 \right\} \, \chi(x) \, dx \,.
\label{Echi}
\ee
Formally, this is the standard static energy of $\phi^4$ theory in terms of
the deformed spatial coordinate $y$, because $dy = -\chi(x) \, dx$, 
except that the endpoints of the $y$-integration may be
non-standard. This energy differs from previous self-dual impurity
models which have eq. (\ref{Impur}) as the corresponding Bogomolny
equation \cite{solvable-imp}.
 
In the usual way, we can complete the square in the integrand, and
obtain
\be
E_{\chi} = -\half\int_{-\infty}^\infty 
\left\{ \frac{1}{\chi(x)}\left(\frac{d\phi}{dx}\right) 
+ (1 - \phi^2) \right\}^2 \, \chi(x) \, dx + \left[ \phi(x) - \frac{1}{3}
\phi^3(x) \right]_{-\infty}^\infty \,.
\ee
The last term depends only on the field topology -- the boundary
data of $\phi$. The energy $E_{\chi}$ is stationary for solutions of 
eq.(\ref{Impur2}), because a change in $\phi$ of order $\varepsilon$ 
changes $E_\chi$ at order $\varepsilon^2$, though it is not guaranteed 
to be a minimum unless $\chi$ is everywhere negative. The energy 
value is $E_{\chi} = \frac{4}{3}$ if $\phi(\infty) = 1$ and 
$E_{\chi} = 0$ if $\phi(\infty) = -1$; it can be zero because the energy 
density is negative in any region where $\chi$ is positive.

It is straightforward to extend the energy function (\ref{Echi}) to
deal with iterated kinks. Define
\be
E = - \half\int_{-\infty}^\infty 
\sum_{n=1}^\infty \mu_n 
\left\{ \frac{1}{\phi_{n-1}^2(x)}\left(\frac{d\phi_n}{dx}\right)^2 
+ (1 - \phi_n^2(x))^2 \right\} \, \phi_{n-1}(x) \, dx \,,
\ee
where $\mu_n$ are fairly arbitrary positive numbers whose sum is finite.
We require that $\phi_n$ has zero derivative at all locations where 
$\phi_{n-1}$ is zero. Completing the square in each term of the sum,
we see that $E$ is stationary when the sequence of iterated kink 
equations is satisfied. 

\section{Summary}

We have introduced a new, iterated equation for kinks in
$\phi^4$ theory. This was motivated by examples of how impurities can
affect a kink. Each equation in the iterated sequence is a first
order, static ODE, whose solution includes one new modulus, the 
constant of integration. In the iterated scheme, the first iteration 
generates a kink from the vacuum. At the second iteration, this kink 
is an impurity which acts like a mirror. It generates a 
kink-antikink configuration, or a positive or negative bump solution 
around the vacuum, depending on the value of the constant of 
integration. The third iteration can produce a kink deformed by a 
variant of the kink's shape mode, and also kink-antikink-kink 
configurations. 

More generally, the $n$th iterate generates an $n$-dimensional moduli 
space of solutions which we propose could be useful for modelling the 
dynamics of $n$ kinks and antikinks. The bump-like configurations 
capture the type of fields that occur dynamically when kinks and antikinks
annihilate, and that are missed in some existing collective coordinate schemes. 
It would be interesting to use the standard $\phi^4$ theory Lagrangian
to calculate the metric (equivalently, the kinetic energy for 
time-varying moduli) and potential energy on these moduli spaces, and 
to study in detail the classical and quantized dynamics of kinks using
these novel collective coordinates.

\renewcommand{\theequation}{A.\arabic{equation}}
\setcounter{equation}{0} 

\section*{Appendix: Iterated polynomials}

An analogy for the iterated kink equation (\ref{Iter}) is the iterated 
equation
\be
\frac{du_n}{dx} = -u_{n-1} \,, \quad n=0,1,2,\dots \,.
\label{Iterpoly}
\ee
Equation (\ref{Iterpoly}) is also the linearisation of eq.(\ref{Iter}) for 
$\phi_n \simeq 0$. We fix $u_{-1}(x) = 0$. Then, generically, $u_0$ is 
a constant, $u_1$ is linear in $x$, $u_2$ is quadratic, and so on. 
$u_{n}(x)$ is a polynomial of degree $n$, so it has at most $n$ real
zeros. A zero can be regarded as analogous to the location of a kink or
antikink, depending on whether $\frac{du_n}{dx}$ is positive or
negative at the zero. Exceptionally, $u_n$ is a polynomial of degree 
$n-k$ if the first non-zero function in the sequence is $u_k$, a 
non-zero constant.

At each iteration, the number of zeros can increase by at most 1, by
Rolle's theorem. If $u_{n-1}$ has $N$ zeros, then $u_n$ has at most $N+1$
zeros, or fewer by a multiple of 2. There can never be 
catching up in the number of zeros. $u_n$ cannot have $n$ zeros 
if any $u_k$, for $k < n$, has fewer than $k$ zeros.

The iteration has fixed points satisfying the equation
\be
\frac{du}{dx} = -u \,,
\ee
whose solutions are 
\be
u(x) = A e^{-x} \,.
\ee
As for the iterated kink equation, a fixed point solution $u(x)$ satisfies
different boundary conditions from any of the sequence of solutions $u_n(x)$.
The fixed point $u(x) = Ae^{-x}$ is the non-uniform limit of the
sequence
\be
u_n(x) = A\left(1 - x + \half x^2 - \frac{1}{6} x^3 + \cdots +
  \frac{(-1)^n}{n!}x^n \right) \,,
\ee
which satisfies eq.(\ref{Iterpoly}).

A 2-cycle satisfies the equations
\be
\frac{du}{dx} = -v \,, \quad \frac{dv}{dx} = -u \,.
\ee
Solutions which are not pure exponentials are of the form
\be
u(x) = -A \cosh x \,, \quad v(x) = A \sinh x \,,
\label{uv2cycle}
\ee
or translates of this. $u$ and $v$ can also be exchanged. The 2-cycle
solution (\ref{2cyclephi}) and (\ref{2cyclepsi}) of the iterated kink 
equation reduces for large $a$ and modest $x$ to the form (\ref{uv2cycle}), 
with $A = \frac{1}{\sqrt{a}}$. See Fig. \ref{phi6figs}.

%%%%%%%%% Acknowledgements %%%%%%%%%%%%%
\section*{Acknowledgements}
%%%%%%%%%%%%%%%%%%%%%%%%%%%%%%%%%%%%%%%%

NSM is grateful to A. Fordy for useful comments. AW thanks 
T. Romanczukiewicz and Ya. Shnir for discussions. This work has 
been partially supported by STFC consolidated grant ST/P000681/1.

\end{document}